\documentclass[conference,10pt]{IEEEtran}
\IEEEoverridecommandlockouts
\usepackage{microtype}                 % use micro-typography (slightly more compact, better to read)
\PassOptionsToPackage{warn}{textcomp}  % to address font issues with \textrightarrow
\usepackage{textcomp}                  % use better special symbols
\usepackage{mathptmx}                  % use matching math font
\usepackage{times}                     % we use Times as the main font
\usepackage[]{cite}                      % needed to automatically sort the references

\usepackage{tabu}                      % only used for the table example
\usepackage{booktabs}                  % only used for the table example
\usepackage{amsmath,amssymb,amsfonts}

\usepackage{algorithmic}
\usepackage{graphicx}
\usepackage{xcolor}
\usepackage[hidelinks]{hyperref}
\usepackage{mathdots}
\usepackage{siunitx}

\usepackage{blkarray, bigstrut}
\usepackage{multirow}
\usepackage{subcaption}
\usepackage{lipsum}

\usepackage{enumitem}
\usepackage{footnote}
\usepackage{color, colortbl}
\usepackage[hidelinks]{hyperref}
\usepackage[T1]{fontenc}

\begin{document}

%% Paper title.

\title{Re-shaping Post-COVID-19 Teaching and Learning: A Blueprint of Virtual-Physical Blended Classrooms in the Metaverse Era}

%% This is how authors are specified in the conference style
%\and Carlos Bermejo\thanks{e-mail: cbf@connect.ust.hk}\\ %
%    \scriptsize{Computational Media \& Arts}\\ %
%    \scriptsize{Hong Kong University of Science and Technology} % 
%\and Lik-Hang Lee\thanks{e-mail: likhang.lee@kaist.ac.kr}\\ %
%    \scriptsize{Augmented Readlity and Media Lab}\\ %
%    \scriptsize{Korea Advanced Institute of Science and Technology}\\ %
%\and Tristan Braud\thanks{e-mail: braudt@ust.hk}\\ %
%    \scriptsize{Computational Media \& Arts}\\ %
%    \scriptsize{Hong Kong University of Science and Technology} %  

%% Author and Affiliation (multiple authors with multiple affiliations)

%\author{Yuyang Wang\thanks{e-mail: yuyangwang@ust.hk}\\ %
%  \scriptsize{Computational Media \& Arts Thrust}\\ %
%    \scriptsize{Hong Kong University of Science and Technology}\\ %
%\and Lik-Hang Lee\thanks{e-mail: likhang.lee@kaist.ac.kr}\\ %
%    \scriptsize{Augmented Reality and Media Lab}\\ %
%    \scriptsize{Korea Advanced Institute of Science and Technology}\\ %
%\and Tristan Braud\thanks{e-mail: braudt@ust.hk}\\ %
%    \scriptsize{Division of Integrative Systems and Design}\\ %
%    \scriptsize{Hong Kong University of Science and Technology} \\%  
%\and Pan Hui, IEEE Fellow\thanks{e-mail: panhui@ust.hk}\\ %
%    \scriptsize{Computational Media \& Arts Thrust}\\ %
%    \scriptsize{Hong Kong University of Science and Technology}\\ %
%}

\author{
\IEEEauthorblockN{Yuyang Wang$^\ast$}\thanks{$^\ast$e-mail:yuyangwang@ust.hk}
\IEEEauthorblockA{\textit{Hong Kong University of Science and Technology (Guangzhou)}\\
\textit{Hong Kong University of Science and Technology}}
\and
\IEEEauthorblockN{Lik-Hang Lee$^\dag$}\thanks{$^\dag$e-mail:likhang.lee@kaist.ac.kr}
\IEEEauthorblockA{\textit{Korea Advanced Institute of Science and Technology}} 
\and
\IEEEauthorblockN{Tristan Braud$^\ddag$}\thanks{$^\ddag$e-mail:braudt@ust.hk}
\IEEEauthorblockA{\textit{Hong Kong University of Science and Technology}}
\and
\IEEEauthorblockN{Pan Hui$^\S$}\thanks{$^\S$e-mail:panhui@ust.hk}
\IEEEauthorblockA{\textit{Hong Kong University of Science and Technology (Guangzhou)}\\
\textit{Hong Kong University of Science and Technology} }
}

\maketitle

\begin{abstract}
During the COVID-19 pandemic, most countries have experienced some form of remote education through video conferencing software platforms. However, these software platforms fail to reduce immersion and replicate the classroom experience. The currently emerging Metaverse addresses many of such limitations by offering blended physical-digital environments. This paper aims to assess how the Metaverse can support and improve e-learning. We first survey the latest applications of blended environments in education and highlight the primary challenges and opportunities. Accordingly, we derive our proposal for a virtual-physical blended classroom configuration that brings students and teachers into a shared educational Metaverse. We focus on the system architecture of the Metaverse classroom to achieve real-time synchronization of a large number of participants and activities across physical (mixed reality classrooms) and virtual (remote VR platform) learning spaces. Our proposal attempts to transform the traditional physical classroom into virtual-physical cyberspace as a new social network of learners and educators connected at an unprecedented scale.
\end{abstract}


%% A teaser figure can be included as follows, but is not recommended since
%% the space is now taken up by a full width abstract.
%\teaser{
%  \includegraphics[width=1.5in]{sample.eps}
%  \caption{Lookit! Lookit!}
%}

\section{Introduction}

% \textcolor{blue}{1. Describe the background of post-covid19 teaching - hybrid online/offline is the new `sexy' trend}

% \textcolor{blue}{2. check any e-learning theory that supports `an evolution' from traditional learning - mobile learning - VR/AR learning - ultimately to our classroom}

Under the complex and changing global political and economic situation and the global pandemic, individual activities and production methods face increasing challenges. For example, lectures have moved to online via Zoom or MS Teams to maintain social distancing. However, the current online educational content is mainly based on flat 2D display, which lacks immersion and engagement compared to traditional teaching in a physical classroom. Students feel hardly focused on the remote lecture. In this situation, the Metaverse appears as a meaningful solution by integrating state-of-the-art technologies such as virtual reality (VR) and augmented reality (AR), artificial intelligence and cloud computing~\cite{lee2021all} to make educational activities to more attractive.

\textbf{Motivations.} The next generation of education, i.e., teaching and learning, should include diversified yet customized learning content and context, increased student interaction and creativity, boosted motivation and engagement~\cite{Erturk2020}. Unfortunately, existing teaching facilities and tools fail to offer these features.  
This work examines existing teaching and learning modalities and proposes a novel teaching system to promote learning experience and performance in the Metaverse era. In this context, we first present a mini-survey on digital education platforms, including computer-mediated learning and teaching via online conferencing tools and VR/AR-based teaching. We highlight the pros and cons of these teaching modalities.  
For instance, Zoom enables synchronous teaching but lacks students engagement~\cite{CHI-21-onlinelearning}.
%For example, teaching through Zoom enables teachers and students to communicate quickly and synchronously. However, it lacks engagement and attraction~\cite{CHI-21-onlinelearning}.
On the other hand, VR/AR provides more engaging teaching activities thanks to 3D visualization technology in semi-immersive or fully immersive environments. However, it requires a high-quality and robust network for synchronizing graphical data~\cite{zhang2020virtual}. Therefore, we put forward a virtual-physical blended classroom, with the ambition of migrating physical classrooms and remote educational activities to the unified yet immersive cyberspace, known as the Metaverse. Our proposed Metaverse classroom contains two physical classrooms at two university campuses and one digital classroom hosted in the edge-cloud synchronized computing devices. The digital classroom enables users (i.e., learners and educators) to appear at a physical lecture from their home campus, in which the digital twins of such users, as avatars, can interact with users from other campuses. %while users from others campuses and online users are represented by avatars. So teachers and students can interact with both online and offline peers.  

%\textcolor{red}{to continue}
%\textbf{Motivations}: The next generation of education, i.e., teaching and learning, should include diversified yet customized learning content and context, increased student interaction and creativity, boosted motivation and engagement~\cite{Erturk2020}. Unfortunately, existing teaching facilities and tools fail to offer these features.  
%This work seeks to examine existing teaching modalities and propose a novel teaching system to promote learning experience and performance in the Metaverse era.

%provide a more engaging and pleasant experience for remote education so that students can learn as efficient as in traditional physical classroom. 

%Also, to optimize teaching interactivity and enhance teaching experience, this auxiliary system will use multimodal perception algorithms to enrich students' and teachers' avatars in the virtual world so that online teaching can also be appealing.

%In this context, this article will propose an virtual-physical blended classroom settings in Metaverse environment. 

\textbf{Review methodology.} First, we search in Google Scholar for the terms ``\textit{VR education}'' and ``\textit{AR education}'', and additional keywords including ``\textit{application}'' and ``\textit{challenge}'' and find that related publications spread among several journals and conferences. Second, we choose papers mainly from celebrated publishers: ACM, IEEE, Springer, Elsevier and Taylor $\&$ Francis. Third, we favour the paper with top citations and from the field of education and VR/AR. Finally, if no proper references can be found in the former steps, we expand our search domain in Google Scholar with other keywords such as ``VR/AR applications'' and filter by investigating the relevance to the educational field. This mini-survey illustrates the paradigm shift in teaching and learning in the post-COVID-19 period. The classroom moves from computer-mediated learning or VR/AR-based applications to a fully virtual-physical blended one.

\textbf{Contributions.} This paper's contributions are twofold, as follows. First, we review the latest development of remote learning and interactive medium like VR/AR. Second, we propose the Metaverse classroom containing two physical classrooms and one cloud-based virtual classroom, and the architecture about how to implement the system.
We accordingly pinpoint the grand challenges of establishing the proposed Metaverse classroom.

%For such reason, many previous studies try to develop adaptive or customized navigation techniques in immersive environments. 

\section{Recent Education Landscape}\label{recent_education}

\begin{figure*}[t]
\centering
\includegraphics[width=0.9\textwidth]{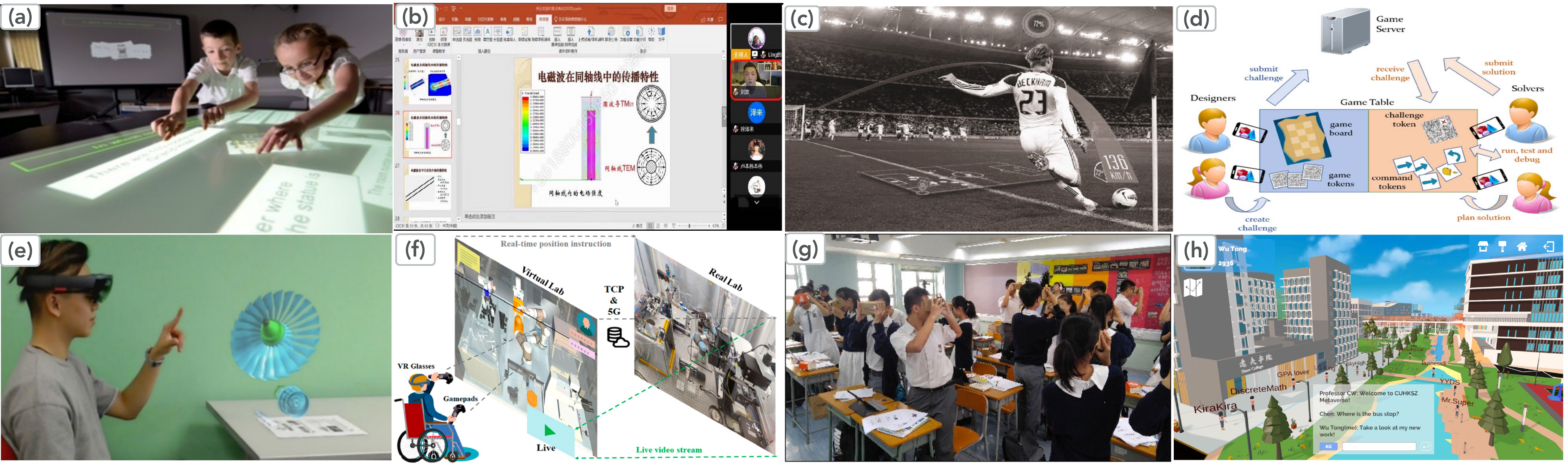}
\caption{Examples of different digital teaching and learning approaches through computer-aided tools and VR/AR technologies. (a) Star trek room to boost children's mathematical abilities with on one desktop~\cite{durham2012}; (b) Teaching via online conferencing tools~\cite{CHI-21-onlinelearning}; (c) Providing additional information during sport training~\cite{soltani2020augmented}; (d) Collaborative learning via ARQuest~\cite{Gardeli2019}; (e) Digital content representation of teaching material~\cite{Tang2020}; (f) Controlling the real lab inside the VR~\cite{Lu2021}; (g) Teaching physical geography with video-based VR~\cite{Jong2020}; (h) Construction of a virtual campus at CUHK SZ~\cite{Duan2021}.}
\label{fig:e-learning} %
\end{figure*}

%\textcolor{red}{The difference between AR and VR}
The recent development of VR/AR provides evidence that immersive elements are no longer a myth in learning context~\cite{Elmqaddem2019}. We highlight some important works related to computer-mediated education and relevant studies of VR/AR, as presented in \autoref{fig:e-learning}. 

%Unfortunately, learners from disadvantaged families still find it hard to overcome the challenges imposed by COVID-19 because they lack stable network connection, powerful computer and digital skills~\cite{onyema2020impact}. 

\textbf{Computer-mediated Education.} 
Since the outbreak of COVID-19, the lack of high-quality learning materials and efficient teaching tools has posed a threat to traditional classroom-based education. Learning remotely through a video conferencing system has provided an opportunity to maintain lecturing and communicating activities. However, the human-computer interaction (HCI) community also identified some challenges arising from remote teaching~\cite{CHI-21-onlinelearning}. For instance, students may find it challenging to pay continuous attention because of multitasking and unexpected interruption. They tend to experience decreased learning efficacy as live streaming learning costs a longer time and reduced engagement/collaboration because the learning environment changes from a community-based public area supported by various university facilities to a private space with limited resources. Researchers proposed several solutions to mitigate the deficiency or weakness for both teachers and students during the live streaming course, e.g., establishing the voice-based conversational agents to engage natural interaction~\cite{Sara-chi20-onlineEDu}, assigning learning tasks to learners and improving motivational beliefs~\cite{zhang2019investigating}, boosting explanation quality and presentation skills~\cite{shoufan2019motivates}.

Durham University developed a multi-touch and multi-user desk to boost children's mathematical skills in a collaborative and interactive way~\cite{higgins2012multi}. Pupils can work together to solve problems and answer questions through some inventive solutions. Compared to traditional paper-based activities, the multi-touch table can improve mathematical flexibility and fluency and promote higher levels of active student engagement.

%With the advent of the COVID-19 pandemic, the insufficiency of the learning content delivery by the traditional education systems has recently come to light. 
%Although online remote learning through video conferencing can serve as an immediate solution in the crisis time, HCI researchers also pinpointed the difficulties of conducting classes remotely~\cite{CHI-21-onlinelearning}, and proposed solutions to alleviate the inconvenience for both learners and instructors~\cite{Sara-chi20-onlineEDu}. However, the challenges and difficulties encountered by early-age learners from low-income families during the crisis of COVID-19 have not been addressed.

\textbf{VR in Education.} VR integrates five essential components: 3D perspective, dynamic rendering, closed-loop interaction, ego-referenced perspective and enhanced sensory feedback~\cite{Wickens1992}. Thanks to rising computing power and the easy access to many affordable head-mounted displays (HMDs), VR has proved to be helpful in various applications such as surgical operations, psychotherapy and STEM education~\cite{Fabris2019}.

%The public mainly believes that such technology is used for gaming and entertainment while limited to other educational or industrial fields. 

%inside-out~(ego-referenced)

%We focus on the application of VR in educational activities. 

The lack of opportunities for practical and hands-on manipulation of objects in the real world is linked to the fact that children are usually poor at intuitive physics~\cite{brelsford1993physics}. Therefore, Brelsford~\cite{brelsford1993physics} designed a physics simulator to give students a better intuition about the physical process inside an immersive environment. One group of students was given a pendulum in controllable length and three balls in varied mass in this simulator. These students can manipulate gravity, mass, location of the balls, air drag, friction, period of the pendulum, force, motion, and starting force of the pendulum to carry out experiments for an hour. Another control group of students was given a lecture in the physical classroom on the same material. An exam was assigned four weeks after the test. The results indicated that those who had learned through the virtual laboratory revealed better retention than those from the lecture-based learning group.

Similarly, Yalow and Snow~\cite{yalow1980individual} reported that offering learners instructional material in spatial-graphical format, instead of verbal words, will improve their immediate comprehension of the material. Therefore, many studies have investigated the benefits of learning in virtual environments. Rega and Fink \cite{rega2014immersive} performed a pioneer study that designed a novel approach to pandemic preparedness and response based on immersive simulation. The authors developed a semester-long course in simulated environments for students in Master of Public Health (MPH). Students are grouped to represent different country health departments during the whole semester, and they can gain incident command training and acquire audio lectures and learning materials related to the imminent pandemic. Despite learning in virtual environments, students still considered this training paradigm more groundbreaking, fascinating and educative. Such work provides us with another solution to combat the current COVID-19 global pandemic~\cite{lieux2021online}. In addition, VR has made significant contributions to other educational cases, e.g., controlling the real chemistry lab by interacting with a virtual lab~\cite{Lu2021} and teaching physical geography through spherical video-based VR immersion~\cite{Jong2020}.

VR can be promising for education and training, but challenges still exist~\cite{Velev2017}. First, VR is generally regarded as a game for entertainment and relaxation. Learners pay more attention to winning in the game, but they are not fully engaged to acquire knowledge and improve critical thinking skills, far from the teaching objective. Second, VR is not perfect from a psychological point of view. For example, the older lectures may prefer the classical teaching approach instead of the digital one compared to younger students. Longstanding immersion in locomotion-dominated VR applications will cause discomfort and sickness symptoms (e.g., nausea, dizziness and disorientation) individually among users, which will weaken users' willingness for using VR devices~\cite{Wang2021}. Third, the educational institution will share the high cost of design and creating the VR resources towards different teaching objectives, and the library department lacks interoperability standards or optimal practices for adopting these VR contents, leading to difficulties to share resources among different institutions and pointless duplicated work~\cite{cook2019challenges}.

\textbf{AR in Education.} 
AR is a 3D technology that boosts the user's perception of the physical work by adding a contextual layer of information~\cite{azuma1997survey}, and it has become a favoured topic in educational research in the last decade~\cite{Ibanez2018}. The popularity of AR technology is because it does not require expensive hardware and complicated equipment such as HMDs. AR can be achieved with low-cost handheld devices such as mobile phones and tablets, enabling AR-based educational settings to be available for most learners. Among many learner types,  K-12 students (primary and secondary students) are the most preferred learners because the outstanding visualization features of AR hold the key to students whose learning performance relies mainly on seeing, hearing or other ways to sense at this age~\cite{Akcayr2017}.

There are many examples of the successful applications of AR in education. Gardeli and Vosinakis~\cite{Gardeli2019} designed the ARQuest application through a collaborative mobile AR game for improving problem-solving and computational skills of primary school students through a gamified activity, which demonstrates the application of AR system in classroom-based interventions. ARQuest supports multi-user interaction, and students can collaborate with high engagement and motivation to solve challenging problems, despite the small screen size. This work can be regarded as a 3D version of the above-mentioned 2D Star trek classroom~\cite{higgins2012multi}. The AR application can benefit learners in sports education and training to learn sports skills, provide additional hints and feedback, stimulate practice, and introduce new rules for creating new sports~\cite{soltani2020augmented}.

However, a broad application of AR in education is still facing some challenges. The most blamed challenge is usability because most AR educational settings are difficult for students to use~\cite{Akcayr2017}. For example, students might find it challenging to use the AR application and perform interactive activities without well-designed interfaces, leading to worse effectiveness~\cite{munoz2014supporting}. In addition, previous work revealed that the group of learners with AR devices required significantly longer training time compared with those without using AR equipment, which might be due to the novelty of the AR technology and users were not familiar with the learning approach~\cite{Gavish2015}. Another issue is that the AR learning environment may increase learners' cognitive workload because of the excessive materials and complicated tasks~\cite{cheng2013affordances}. Some technical issues can also weaken learners' expectations of AR technology. For example, location-based AR applications rely heavily on the GPS signal to determine the position and orientation, while low sensitivity in trigger recognition frequently appears as an issue~\cite{cheng2013affordances}.

%\cite{gavish2015evaluating}

\section{A vision of Metaverse Classrooms} %Prototype}

Traditional teaching activities depend on verbal and nonverbal interactions between teachers and students to formulate and enrich cultural norms, behaviors, practices, and beliefs, but meeting on an online platform interrupts such a process and changes individuals‘ communication habits~\cite{Greenan2021}. On the other hand, current VR/AR education allows 3D visualization but fails to provide remote access. Therefore, we expect to overcome the defects of current education methods and therefore propose the virtual-physical blended Metaverse classroom, which can boost student engagement via more sense of presence, improve learning efficiency via 3D visualization, and create supportive connection tools via immersive interaction during and after class meetings. The Metavese classroom room can take advantage of the online and VR/AR education, therefore avoiding some disadvantages of each.

Greenan~\cite{Greenan2021} argued that social presence and self-disclosure are two crucial aspects of virtual education. Garrison et al.~\cite{garrison1999critical} explained the social presence as socio-emotional support and interaction and the individual's ability to project themselves socially and emotionally with their entire personality through a communication tool, which will be impacted by conversions, activities, collaboration, familiarity and motivation among participants. On the other hand, during virtual education, self-disclosure could lower the feeling of unreliability and ambiguity during communication and promote intimacy and positive relationships~\cite{Greenan2021}, which reduce the mental distance between individuals and increase trust between teachers and students~\cite{song2019know}. Thus, with the increasing popularity of AR and VR education applications (\autoref{recent_education}), establishing the Metaverse classroom can promote the aforementioned benefits at scale.

\subsection{Towards the Metaverse classroom }

Due to the global pandemic, it is limited for teachers and students access to classrooms and laboratories. Online conferencing tools enable continuity of education, making lectures and tutorials available virtually to anyone. Coincidentally, the pandemic shed light on the feasibility of e-learning in long duration and hence the user acceptability to the e-learning at scale. However, engagement is an essential component of education, and passive learning via video conferencing fails to reinforce the engagement among class participants. One way to tackle this challenge, highlighted during the COVID-19 pandemic, is to introduce interactive and real-time elements to educators and learners. The advanced development of VR/AR is characterized by immersive visual stimuli and real-time tracking. Also, the related hardware and multimedia technologies can support the Metaverse as an affordable learning platform for learners and educators on the globe. With such an unprecedented opportunity, we can employ the Metaverse as a new social platform for both learners and educators, as nowadays VR/AR devices are getting mature, which serve as an efficient tool for teaching and learning.  %simultaneously offer immersive visual stimuli and real-time tracking.
 %do not require a tedious setup, 

%One way to tackle this challenge being highlighted during the COVID-19 panademic is to introduce interactive and real-time elements to educators and learners. %, such as polls, in teaching. 
%Another approach is to employ Metaverse technologies. Modern VR/AR/MR devices do not require a tedious setup, provide immersive visual stimuli, real-time tracking, and can be efficiently used as a tool for teaching and learning. 

% HKUST 

We propose an educational platform that supports the teaching and learning experience with the Metaverse, with the following features:  
(i) learning assessment in the Metaverse for the courses, (ii) interaction with presentations in the Metaverse, and (iii) teaching experience with augmentation that benefits 
from visualizing knowledge and
from 3D virtual entities. The platform aims to bring 
more engaging and interactive tutorials and mixed learning. 
Remarkably, the virtual-physical blended classroom in the Metaverse makes the educators and learners situated in intuitive yet user-context (to both learners and educators) environments. Thus, the Metaverse classroom provides effective communication media among participants from various campuses. At the same time, the changeover in such a virtual-physical environment brings new values to the class participants, in terms of learning effectiveness, senses of (virtual) presences, interactive learning experience, etc. Although such a Metaverse classroom can connect to numerous possible usage scenarios, we specifically highlight several class participants' interactions, as follows.

\textbf{Gamified Learning and Task-based Modules.} Promoting gamification is obvious, which facilitates active class dynamics %movement 
by designing digital ``breakouts" for teams of students or letting students collaborate to create their own.  

\textbf{Learner Collaborations.} Challenging students to work in teams to solve a riddle or puzzle or bring the surreal scenes, i.e., augmented and virtual realities, %magic of virtual reality 
to the classroom. Additionally, incorporating a 360-degree video scene. 

\textbf{Learner-driven Activities.} Empowering students to create ``choose your own adventure"-style stories or presentations to share their contents (learning outcomes, opinions, a speech, \textit{etc}.) with the Metaverse community. %Additionally, the Metaverse classrom encourages s or create a professional development experience that will have teachers thinking like kids again! 

\textbf{Saving Instructors' Time.}	With unlimited possibilities for creating experiences and new features being added continuously, instructors would become more willing to spend their time on designing new curricula, creating diversified and even personalized activities for the sake of improving their students' skills and knowledge. % can design or empower students to create activities that teach various skills. 

%Remarkably, many students cannot access labs due to the COVID-19.
\textbf{Access to Limited/restricted Equipment.} As people can access the Metaverse classroom anytime and anywhere, the virtual-physical presence of class participants could achieve ubiquitous learning yet real-time access to the lab resource (e.g., a virtual lab as the digital twin) as well as other limited/restricted resources (e.g., testing Uranium in the Metaverse). %Through its ability to take people anywhere, virtual reality can be a powerful resource for students who otherwise would not have access to the lab time they need to complete their degrees.

\begin{figure}[!ht]
\centering
\includegraphics[width=0.48\textwidth]{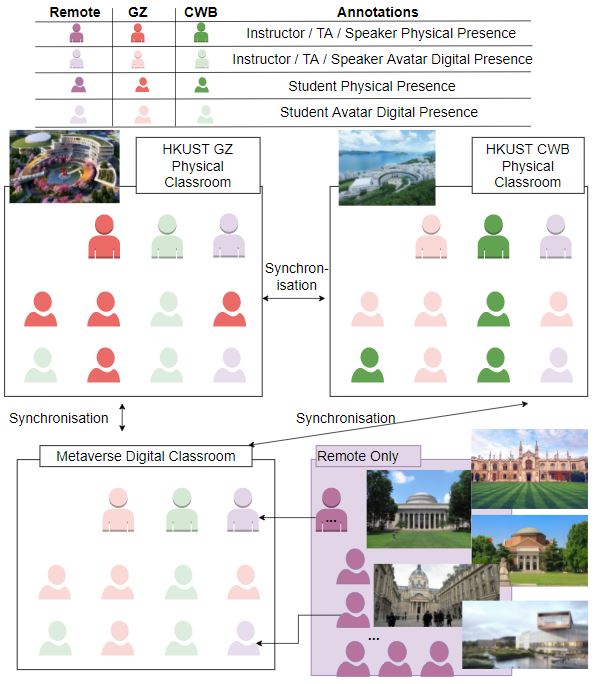}
\caption{The MR Metaverse classroom allows students, instructors, and speakers located physically in HKUST GZ campus, HKUST CWB campus, or online to interact within the same activity (lecture, tutorial, seminar). Through MR, the virtual classroom enables users to attend a physical lecture on their home campus, while users from other campus and online users are represented through avatars.}
\label{fig:MR_room} %
\end{figure}

We consider the Metaverse classroom consists of both physical and virtual classrooms. As shown in \autoref{fig:MR_room}, learners and instructors are situated in classrooms spreading over multiple campuses, while the virtual classroom bridge the physical class participants together. The physical participants of one location can meet the virtual participants and the physical participants of other locations. We outline the above concept with a unit case. The participants are situated in two physical classrooms and one virtual classroom spreading over two campuses (Hong Kong Clear Water Bay, CWB and Guangzhou, GZ) in the following paragraphs. These three classrooms are synchronized so that the intervention of a participant in any of these classrooms will be visible to the attendants in the other two classrooms through his or her avatar representation (i.e., the digital twins of class participants). It is important to note that the number of physical classrooms is expandable, i.e., more than two physical classrooms can access the Metaverse classroom. %, in addition to the virtual class participants With HKUST spreading over two campuses, this project aims to develop an XR Metaverse classroom to enable simultaneous physical lectures on both campuses while allowing online users to connect remotely. The Metaverse classroom will leverage the full extent of XR technology to regroup three classrooms, as follows:

\textbf{Physical Metaverse Classroom in HKUST GZ and CWB Campuses.} These two classrooms are linked together through a mixed reality (MR) Metaverse platform that enables classes to be shared between the two campuses. Each classroom is equipped with a set of sensors to track the participants and replicate their presence through digital avatars on another campus' classroom(s). These digital avatars are displayed in MR through projectors or headsets. As such, it is possible to seamlessly conduct a wide range of activities, ranging from talks and lectures to group projects involving students and staff from both campuses.

\textbf{Digital Metaverse Classroom Online in VR.} A significant portion of class attendants can access the learning context remotely, i.e., without any physical presence, which is similar to nowadays situation of remote learning via Zoom. Nonetheless, the remote participants can virtually meet the physical participants from HKUST GZ and CWB campuses. 
%We expect some users to also attend classes fully remotely, without any physical presence. 
These users perhaps are either HKUST students who cannot attend the physical lecture due to unexpected circumstances (sickness, travel restriction, etc.) or learners outside of HKUST who audit the course. 
These users can connect to a third, fully virtual classroom through a VR headset or their computers. 
The HKUST students' presence can be represented through digital avatars displayed in both MR classrooms, while guest avatars enable outside users to participate in the class (e.g., guest speakers). 
For instance, students from KAIST, represented by avatars, can meet HKUST students in the Metaverse Digital Classroom, as well as other remote class participants from MIT and Cambridge (the lower half in~\autoref{fig:MR_room}).

\subsection{Architecture of the Metaverse classroom}

\begin{figure}[!htp]
\centering
\includegraphics[width=0.499\textwidth]{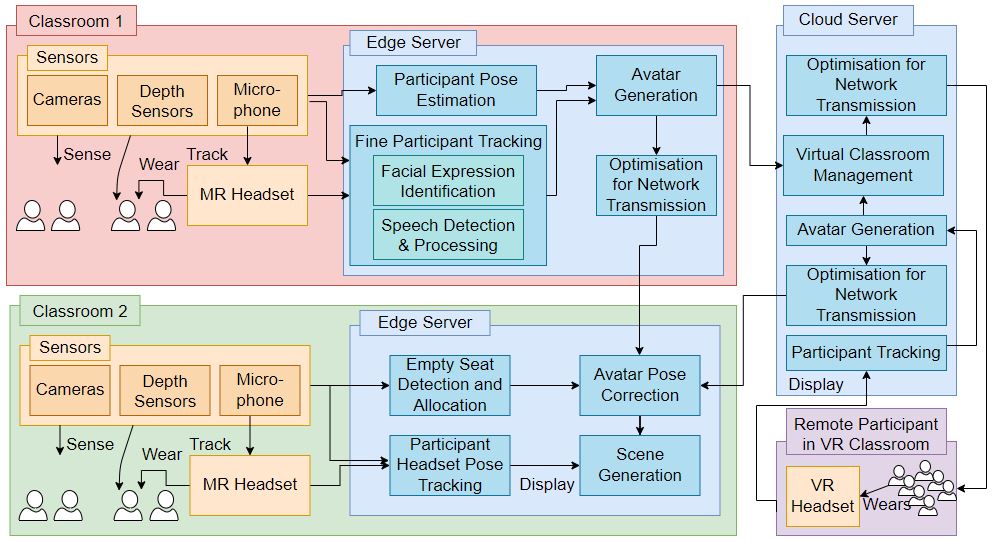}
\caption{General system architecture to handle synchronization between two MR Metaverse classrooms and a VR Metaverse classroom.}
\label{fig:architecture} %
\end{figure}

Following the unit case in previous paragraphs, \autoref{fig:architecture} depicts the system architecture for replicating physical participants situated in the physical classroom (Classrooms 1 and 2), who can interact with remote participants in virtual cyberspace. %in Classroom 1 and remote participants in the mixed reality of Classroom 2.
The participants in the physical classroom 1 wear MR headsets that can track their locations and other features, such as facial expressions. Meanwhile, the physical classroom is equipped with %embeds 
non-intrusive sensors that can estimate the exact pose of the participants. The data from the headsets and the classroom sensors are transmitted through WiFi (headset) or wired network (sensors) to the edge server that aggregates the data to estimate the pose and facial expression of the participants. 

The server then generates the avatar and their interaction traces accordingly, and packages them via the real-time transmission link to both the edge server of Classroom 2 and the cloud server of the VR classroom. 
The edge server in Classroom 2 identifies the vacant seats to display virtual avatars in the MR classroom. Upon the reception of the digital information, it corrects the pose to match the new position of the avatar and generates the scene to display to the users through the lens of their MR headsets. Similarly, the cloud server arranges the avatars of all users within an entirely virtual VR classroom and transmits the results back to the remote users.
Both classrooms can rely on their own independent WiFi infrastructure to accelerate the data transmission between the headsets and the edge servers and minimize the end-to-end latency for replicating the participants’ poses and expressions as digital avatars on the other Metaverse classrooms, whether MR or VR.

\subsection{Challenges for developing the Metaverse classroom}

The Metaverse classroom will be a more student-centered, collaborative and innovative platform for educational activities. Despite being a promising teaching platform, the proposed architecture must be meticulously investigated and designed to minimize some side effects and maximize its potential. Some challenges need addressing during the implementation for keeping the learning engagement and efficiency, and creating teaching aids inside the Metaverse classroom. The overcome of these troubles will also help us solve issues (e.g., usability, cybersickness, and high cost) inherited from VR/AR technology because most problems arise from interaction and content design, systems and networking stability.

\textbf{User Interactivity and Perception.} Learners and educators are the key actors in the Metaverse classroom. Maintaining the bandwidth of user interactivity, regardless of output and inputs, is crucial to user retention. Nonetheless, first, the user inputs on mobile MR and VR headsets are far from satisfaction, resulting in low throughput rates in general~\cite{Lee2022TowardsAR}, which could hinder the users' expression, i.e., converting one's intentions into resultant outcomes in virtual-physical blended environments. Additionally, current input methods of headsets are primarily speech recognition and simple hand gestures~\cite{Lee2020UbiPointTN}. As a result, users can only deal with enriched educational contents/contexts at a degraded quality. On the other hand, the output channels, known as feedback cues, are limited, albeit the headsets can sufficiently provide visuals of digital entities. However, such displays' limited Field-of-View (FOV) potentially deteriorates communication efficacy among users in the Metaverse classroom~\cite{Lee2020FromST}. Specifically, partial view of body gestures, heavily relying on constant visual attention, due to limited FOV, can lead to distorted communication outcomes. Thus, multi-modal feedback cues (e.g., haptics) become necessary to maintain the granularity of user communication. Additionally, haptic feedback is essential to delivering high levels of presence and realism, but current networking constraints create delayed feedback and damage user experiences~\cite{Bermejo2021ExploringBD}. 
More importantly, presence and realism can influence the sense of ``being together'' in the Metaverse classroom. 

\textbf{Navigation and Cybersickness.} 
%start here
Navigation is the most fundamental interaction and is the primary task when users move inside the 3D virtual environment. For example, students need to move from one position to another and adapt their viewpoint in the virtual classroom to communicate with their learning peers. In physical environments, humans can navigate easily and gain consistent multisensory feedback from through from visual, proprioceptive and vestibular systems. In contrast, in immersive virtual environments, they cannot walk naturally due to the limitation of Metaverse devices and small workspace. They have to be constrained into a limited physical space, implying impossible matches between actual and virtual walking, and thus leading to unnatural sensory feedback. According to the sensory conflict theory~\cite{oman1990motion}, the mismatched visual and vestibular information will lead users to experience cybersickness with symptoms, such as fatigue, headache, nausea, disorientation, etc. As there exists a significant correlation between the sickness level and user discomfort~\cite{somrak2019estimating}, failure in mitigating cybersickness would threaten the success and acceptance of the Metaverse classroom. Several technical settings are responsible for the occurrence of cybersickness, such as latency, FOV, low frame rates, inappropriate adjustment of navigation parameters~\cite{Chardonnet2020}. As the user susceptibility to cybersickness is individually different, the Metaverse classroom would consider to ease the severity of cybersickness by involving individual factors such as gender, gaming experience, age, ethnic origin, \textit{etc}~\cite{Wang2021fuzzy}. 

%~\cite{kennedy1993simulator}

% For example, users would visually perceive moving objects while their body is still in position.

%The Metaverse system can acquire users' physiological signals, predict the individual sickness severity, and adapt the learning content and duration accordingly. 

%~\cite{Kim2019}
%In addition to aforementioned factors, an ideal Metaverse classroom should be customized to students' learning habits, interests, response by improving 

\textbf{Systems and Networking.} 
%start here
Developing such a classroom raises significant challenges in synchronization. %of a large number of entities within a single digital space. %The Metaverse classroom combines elements of live video transmission, massively multi-user 3D worlds, computation offloading, and eventually distributed rendering. 
The MR environment requires transmitting large amounts of data to finely synchronize the users' actions. Although these data account for less traffic than live video streaming, users' actions need to be synchronized in real-time to enable seamless interaction. As such, latency is a primary challenge. In highly interactive applications, users start to notice latency above 100\,ms. Besides, a latency below 100\,ms still affects user performance despite less noticeable~\cite{claypool}. Latency will therefore be a primary concern in such a classroom. Another major challenge lies in sharing the real-time course with thousands of remote users scattered worldwide. Although some works strive to address massively multi-users systems~\cite{donkervliet2020towards}, they do not address the stringent latency that may happen when users located either far away, or on a poorly interconnected network ( due to peering agreements or firewalls) present a round-trip latency in the order of the hundreds of milliseconds. Most gaming platforms solve this issue by setting up regional servers. 
%However, the universality of education complicates this task, as a course taught in a certain geographical location on the Metaverse should be accessible to anybody with internet connection. 
Another major concern is the fine rendering of the digital avatars of physical participants. Owing to the pervasive sensing capabilities of the physical MR classroom, it will be possible to rebuild highly accurate representations of the physical participants as sophisticated avatars. However, these avatars may be too complex to render with WebGL and lightweight VR headsets. As such, it may be necessary to leverage servers (cloud and edge) to pre-render some elements of the digital scene. One solution would be to render a low-quality version of the models on-device and merge the rendered frame with high-quality frames rendered in the cloud~\cite{lee2015outatime}. Finally, many courses may rely on video transmission, whether of the instructor, digital artefacts (e.g., slides), or physical objects in the classroom (e.g., whiteboard). These video frames need to be transmitted in real-time to match both the avatars' actions and the related audio transmission. Maximizing video quality while minimizing latency to an imperceptible level has been a challenge in the cloud gaming community, and leveraging joint source coding and forward error correction at the application level are promising solutions~\cite{alhilal2022nebula}.

%A high video quality (high resolution with few artifacts) is also necessary to deliver information with high legibility. 

%

\textbf{Content Democratization and Privacy.} 
%start here
The Metaverse encourages participants to contribute content in the virtual-physical blended cyberspace~\cite{Lee2021WhenCM}. Furthermore, regardless of learners and educators, class participants in the proposed classroom are expected to contribute learning content in various educational contexts. NFTs and well-design economics models are the keys to the sustainability of user contributions that expect credits and rewards. Finally, as the newly created content will remain in the classroom cyberspace, we have to consider the appropriateness of content overlays under the privacy-preserving perspective~\cite{Kumar2021TheophanyMS}. Improper augmentation of contents in the Metaverse can pose privacy threats and perhaps risks of copyright infringement.

\section{Conclusion}

% COVID-19 has accelerated the development of the Metaverse classroom. 

%The pandemic catalyzes the emergence of immersive environments that potentially impact how we work, play, and even learn and teach \cite{wijesooriya2020covid}. 

The pandemic catalyzes the emergence of immersive environments that potentially impact how we work, play, and even learn and teach. We foresee that the Metaverse classroom will bring more learner-centric, collaborative, and innovative elements to the future classroom. Our work serves as a groundwork for the digital transformation of learning and teaching, from sole digital medium to virtual-physical blended one.
Despite the Metaverse classroom being a promising learning platform for students, extra efforts are required to meticulously investigate system performance and user-centric evaluation and collect valuable feedback from actual trials, i.e., teaching activities. Therefore, we call for joint research efforts to actualize the Metaverse classroom, and particularly, we have to address several challenging aspects including user-centric, system and networking issues. 

%teaching tool for students, the development of Metaverse classroom encounters several challenges. 

%Future work to promote the development of Metaverse classroom can involve the following aspects. For example, the feasibility and cost effectiveness of the Metaverse classroom need to be validated based on joint teaching activities locally, nationally and internationally. 

%\textbf{Disadvantaged students who doesn't have a good pc and network}

%\section{Conclusion}\label{conclusion}

% How to report mann-whitney u : https://stats.stackexchange.com/questions/87403/how-do-you-report-a-mann-whitney-test

%"Median latencies in groups E and C were 153 and 247 ms; the distributions in the two groups differed significantly (Mann–Whitney U = 10.5, n1 = n2 = 8, P < 0.05 two-tailed)."

% \acknowledgments{...}

\bibliographystyle{IEEEtran}

\bibliography{reference.bib}

\begin{thebibliography}{10}
\providecommand{\url}[1]{#1}
\csname url@samestyle\endcsname
\providecommand{\newblock}{\relax}
\providecommand{\bibinfo}[2]{#2}
\providecommand{\BIBentrySTDinterwordspacing}{\spaceskip=0pt\relax}
\providecommand{\BIBentryALTinterwordstretchfactor}{4}
\providecommand{\BIBentryALTinterwordspacing}{\spaceskip=\fontdimen2\font plus
\BIBentryALTinterwordstretchfactor\fontdimen3\font minus
  \fontdimen4\font\relax}
\providecommand{\BIBforeignlanguage}[2]{{%
\expandafter\ifx\csname l@#1\endcsname\relax
\typeout{** WARNING: IEEEtran.bst: No hyphenation pattern has been}%
\typeout{** loaded for the language `#1'. Using the pattern for}%
\typeout{** the default language instead.}%
\else
\language=\csname l@#1\endcsname
\fi
#2}}
\providecommand{\BIBdecl}{\relax}
\BIBdecl

\bibitem{lee2021all}
L.-H. Lee, T.~Braud \emph{et~al.}, ``All one needs to know about metaverse: A
  complete survey on technological singularity, virtual ecosystem, and research
  agenda,'' \emph{arXiv:2110.05352}, 2021.

\bibitem{Erturk2020}
``{The expanding role of immersive media in education},'' \emph{Proc. of the
  14th IADIS Inter. Conf. e-Learning 2020, EL 2020 - Part of the 14th Multi
  Conf. on MCCSIS 2020}, pp. 191--194, 2020.

\bibitem{CHI-21-onlinelearning}
Z.~Chen \emph{et~al.}, ``Learning from home: A mixed-methods analysis of live
  streaming based remote education experience in chinese colleges during the
  covid-19 pandemic,'' in \emph{Proc. of the 2021 CHI Conf. on Human Factors in
  Comp. Sys.}, ser. CHI '21.\hskip 1em plus 0.5em minus 0.4em\relax NY, USA:
  ACM, 2021.

\bibitem{zhang2020virtual}
Y.~Zhang \emph{et~al.}, ``Virtual reality applications for the built
  environment: Research trends and opportunities,'' \emph{Automation in
  Construction}, vol. 118, p. 103311, 2020.

\bibitem{Elmqaddem2019}
N.~Elmqaddem, ``{Augmented Reality and Virtual Reality in Education. Myth or
  Reality?}'' \emph{International J. of Emerging Technologies in Learning
  (iJET)}, vol.~14, no.~03, p. 234, feb 2019.

\bibitem{durham2012}
``Star trek classroom: the next generation of school desks,'' Durham University
  News, Nov 2012, https://www.dur.ac.uk/news/newsitem/?itemno=15991.

\bibitem{soltani2020augmented}
P.~Soltani and A.~H. Morice, ``Augmented reality tools for sports education and
  training,'' \emph{Computers \& Education}, vol. 155, p. 103923, 2020.

\bibitem{Gardeli2019}
A.~Gardeli and S.~Vosinakis, ``{ARQuest: A tangible augmented reality approach
  to developing computational thinking skills},'' \emph{2019 11th Inter. Conf.
  on Virtual Worlds and Games for Serious Applications, VS-Games 2019 - Proc.},
  p. 1DUUMY, 2019.

\bibitem{Tang2020}
Y.~M. Tang, K.~M. Au, H.~C. Lau, G.~T. Ho, and C.~H. Wu, ``{Evaluating the
  effectiveness of learning design with mixed reality (MR) in higher
  education},'' \emph{Virtual Reality}, vol.~24, no.~4, pp. 797--807, 2020.

\bibitem{Lu2021}
Y.~Lu, Y.~Xu, and X.~Zhu, ``{Designing and Implementing $VR^2E^2C$, a Virtual
  Reality Remote Education for Experimental Chemistry System},'' \emph{Journal
  of Chemical Education}, vol.~98, no.~8, pp. 2720--2725, aug 2021.

\bibitem{Jong2020}
M.~S.~Y. Jong, C.~C. Tsai, H.~Xie, and F.~{Kwan-Kit Wong}, ``{Integrating
  interactive learner-immersed video-based virtual reality into learning and
  teaching of physical geography},'' \emph{British Journal of Educational
  Technology}, 2020.

\bibitem{Duan2021}
H.~Duan, J.~Li, S.~Fan, Z.~Lin, X.~Wu, and W.~Cai, ``{Metaverse for Social
  Good: A University Campus Prototype},'' in \emph{Proceedings of the 29th ACM
  International Conference on Multimedia}.\hskip 1em plus 0.5em minus
  0.4em\relax New York, NY, USA: ACM, oct 2021, pp. 153--161.

\bibitem{Sara-chi20-onlineEDu}
R.~Winkler \emph{et~al.}, \emph{Sara, the Lecturer: Improving Learning in
  Online Education with a Scaffolding-Based Conversational Agent}.\hskip 1em
  plus 0.5em minus 0.4em\relax NY, USA: ACM, 2020, p. 1–14.

\bibitem{zhang2019investigating}
S.~Zhang and Q.~Liu, ``Investigating the relationships among teachers’
  motivational beliefs, motivational regulation, and their learning engagement
  in online professional learning communities,'' \emph{Computers \& Education},
  vol. 134, pp. 145--155, 2019.

\bibitem{shoufan2019motivates}
A.~Shoufan, ``What motivates university students to like or dislike an
  educational online video? a sentimental framework,'' \emph{Computers \&
  education}, vol. 134, pp. 132--144, 2019.

\bibitem{higgins2012multi}
S.~Higgins \emph{et~al.}, ``Multi-touch tables and collaborative learning,''
  \emph{British J. of Educational Technology}, vol.~43, no.~6, pp. 1041--1054,
  2012.

\bibitem{Wickens1992}
C.~D. Wickens, ``{Virtual reality and education},'' \emph{IEEE International
  Conference on Systems, Man and Cybernetics}, 1992.

\bibitem{Fabris2019}
C.~P. Fabris \emph{et~al.}, ``{Virtual reality in higher education},''
  \emph{Inter. J. of Innovation in Science and Mathematics Education}, vol.~27,
  no.~8, pp. 69--80, 2019.

\bibitem{brelsford1993physics}
J.~W. Brelsford, ``{Physics Education in a Virtual Environment},'' \emph{Proc.
  of the Human Factors and Ergonomics Society Annual Meeting}, vol.~37, no.~18,
  pp. 1286--1290, oct 1993.

\bibitem{yalow1980individual}
E.~Yalow and R.~E. Snow, ``Individual differences in learning from verbal and
  figural materials.'' Stanford Univ Calif School of Education, Tech. Rep.,
  1980.

\bibitem{rega2014immersive}
P.~P. Rega and B.~N. Fink, ``Immersive simulation education: a novel approach
  to pandemic preparedness and response,'' \emph{Public Health Nursing},
  vol.~31, no.~2, pp. 167--174, 2014.

\bibitem{lieux2021online}
M.~Lieux \emph{et~al.}, ``Online conferencing software in radiology: Recent
  trends and utility,'' \emph{Clinical Imaging}, vol.~76, pp. 116--122, 2021.

\bibitem{Velev2017}
D.~Velev and P.~Zlateva, ``{Virtual Reality Challenges in Education and
  Training},'' \emph{International J. of Learning}, vol.~3, no.~1, pp. 33--37,
  2017.

\bibitem{Wang2021}
Y.~Wang \emph{et~al.}, ``{Development of a speed protector to optimize user
  experience in 3D virtual environments},'' \emph{Inter. J. of Human-Computer
  Studies}, vol. 147, p. 102578, dec 2021.

\bibitem{cook2019challenges}
M.~Cook \emph{et~al.}, ``{Challenges and strategies for educational virtual
  reality: Results of an expert-led forum on 3D/VR technologies across academic
  institutions},'' \emph{Information Technology and Libraries}, vol.~38, no.~4,
  pp. 25--48, 2019.

\bibitem{azuma1997survey}
R.~T. Azuma, ``A survey of augmented reality,'' \emph{Presence: teleoperators
  \& virtual environments}, vol.~6, no.~4, pp. 355--385, 1997.

\bibitem{Ibanez2018}
M.-B. Ib{\'{a}}{\~{n}}ez and C.~Delgado-Kloos, ``{Augmented reality for STEM
  learning: A systematic review},'' \emph{Computers {\&} Education}, vol. 123,
  pp. 109--123, aug 2018.

\bibitem{Akcayr2017}
M.~Ak{\c{c}}ayır and G.~Ak{\c{c}}ayır, ``{Advantages and challenges
  associated with augmented reality for education: A systematic review of the
  literature},'' \emph{Educational Research Review}, vol.~20, pp. 1--11, 2017.

\bibitem{munoz2014supporting}
J.~A. Munoz-Cristobal \emph{et~al.}, ``Supporting teacher orchestration in
  ubiquitous learning environments: A study in primary education,'' \emph{IEEE
  Trans. on Learning Technologies}, vol.~8, no.~1, pp. 83--97, 2014.

\bibitem{Gavish2015}
N.~Gavish \emph{et~al.}, ``Evaluating virtual reality and augmented reality
  training for industrial maintenance and assembly tasks,'' \emph{Interactive
  Learning Environments}, vol.~23, no.~6, pp. 778--798, 2015.

\bibitem{cheng2013affordances}
K.-H. Cheng and C.-C. Tsai, ``Affordances of augmented reality in science
  learning: Suggestions for future research,'' \emph{J. of science education
  and technology}, vol.~22, no.~4, pp. 449--462, 2013.

\bibitem{Greenan2021}
K.~A. Greenan, ``{The Influence of Virtual Education on Classroom Culture},''
  \emph{Frontiers in Communication}, vol.~6, no. March, pp. 10--13, mar 2021.

\bibitem{garrison1999critical}
D.~R. Garrison, T.~Anderson, and W.~Archer, ``Critical inquiry in a text-based
  environment: Computer conferencing in higher education,'' \emph{The internet
  and higher education}, vol.~2, no. 2-3, pp. 87--105, 1999.

\bibitem{song2019know}
H.~Song, J.~Kim, and N.~Park, ``I know my professor: Teacher self-disclosure in
  online education and a mediating role of social presence,'' \emph{Inter. J.
  of Human--Computer Interaction}, vol.~35, no.~6, pp. 448--455, 2019.

\bibitem{Lee2022TowardsAR}
L.-H. Lee \emph{et~al.}, ``Towards augmented reality driven human-city
  interaction: Current research on mobile headsets and future challenges,''
  \emph{ACM Computing Surveys (CSUR)}, vol.~54, pp. 1 -- 38, 2022.

\bibitem{Lee2020UbiPointTN}
------, ``Ubipoint: towards non-intrusive mid-air interaction for hardware
  constrained smart glasses,'' \emph{Proc. of the 11th ACM Multimedia Sys.
  Conf.}, 2020.

\bibitem{Lee2020FromST}
------, ``From seen to unseen: Designing keyboard-less interfaces for text
  entry on the constrained screen real estate of augmented reality headsets,''
  \emph{Pervasive Mob. Comput.}, vol.~64, p. 101148, 2020.

\bibitem{Bermejo2021ExploringBD}
C.~Bermejo \emph{et~al.}, ``Exploring button designs for mid-air interaction in
  virtual reality: A hexa-metric evaluation of key representations and
  multi-modal cues,'' \emph{Proc. of the ACM on Human-Computer Interaction},
  vol.~5, pp. 1 -- 26, 2021.

\bibitem{oman1990motion}
C.~M. Oman, ``Motion sickness: a synthesis and evaluation of the sensory
  conflict theory,'' \emph{Canadian J. of physiology and pharmacology},
  vol.~68, no.~2, pp. 294--303, 1990.

\bibitem{somrak2019estimating}
A.~Somrak \emph{et~al.}, ``Estimating vr sickness and user experience using
  different hmd technologies: An evaluation study,'' \emph{Future Generation
  Comp. Sys.}, vol.~94, pp. 302--316, 2019.

\bibitem{Chardonnet2020}
J.-R. Chardonnet, M.~A. Mirzaei, and F.~Merienne, ``Influence of navigation
  parameters on cybersickness in virtual reality,'' \emph{Virtual Reality},
  vol.~25, no.~3, pp. 565--574, 2021.

\bibitem{Wang2021fuzzy}
Y.~Wang \emph{et~al.}, ``{Using Fuzzy Logic to Involve Individual Differences
  for Predicting Cybersickness during VR Navigation},'' in \emph{2021 IEEE
  VR}.\hskip 1em plus 0.5em minus 0.4em\relax Lisbon, Portugal: IEEE, mar 2021,
  pp. 373--381.

\bibitem{claypool}
M.~Claypool and K.~Claypool, ``Latency and player actions in online games,''
  \emph{Commun. ACM}, 2006.

\bibitem{donkervliet2020towards}
J.~Donkervliet, A.~Trivedi, and A.~Iosup, ``Towards supporting millions of
  users in modifiable virtual environments by redesigning
  $\{$Minecraft-Like$\}$ games as serverless systems,'' in \emph{12th USENIX
  Workshop on Hot Topics in Cloud Computing (HotCloud 20)}, 2020.

\bibitem{lee2015outatime}
K.~Lee \emph{et~al.}, ``Outatime: Using speculation to enable low-latency
  continuous interaction for mobile cloud gaming,'' in \emph{Proceedings of the
  13th Annual International Conference on Mobile Systems, Applications, and
  Services}, 2015.

\bibitem{alhilal2022nebula}
A.~Alhilal \emph{et~al.}, ``Nebula: Reliable low-latency video transmission for
  mobile cloud gaming,'' \emph{arXiv:2201.07738}, 2022.

\bibitem{Lee2021WhenCM}
L.-H. Lee \emph{et~al.}, ``When creators meet the metaverse: A survey on
  computational arts,'' \emph{ArXiv}, vol. abs/2111.13486, 2021.

\bibitem{Kumar2021TheophanyMS}
A.~Kumar \emph{et~al.}, ``Theophany: Multimodal speech augmentation in
  instantaneous privacy channels,'' \emph{Proc. of the 29th ACM Inter. Conf. on
  Multimedia}, 2021.

\end{thebibliography}

\end{document}